\documentclass[aps,prb,reprint,floatfix,nobibnotes]{revtex4-1}
\pdfoutput=1
\usepackage{mathtools,amssymb}
\usepackage{units,siunitx}
\usepackage{graphicx}
\usepackage[version=3]{mhchem}

\begin{document}

\title{Investigation of the Spatially Dependent Charge Collection Probability in CuInS\textsubscript{2}/ZnO Colloidal Nanocrystal Solar Cells}

\author{Dorothea Scheunemann}
\email{dorothea.scheunemann@uol.de}
\author{Sebastian Wilken}
\email{sebastian.wilken@uol.de}
\author{J\"urgen Parisi}
\author{Holger Borchert}
\affiliation{Institute of Physics, Energy and Semiconductor Research Laboratory, Carl von Ossietzky University of Oldenburg, 26111 Oldenburg, Germany}

\begin{abstract}
Solar cells with a heterojunction between colloidal \ce{CuInS2} and ZnO nanocrystals are an innovative concept in solution-processed photovoltaics, but the conversion efficiency cannot compete yet with devices employing lead chalcogenide quantum dots. Here, we present a detailed study on the charge collection in \ce{CuInS2}/ZnO solar cells. An inverted device architecture was utilized, in which the ZnO played an additional role as optical spacer layer. Variations of the ZnO thickness were exploited to create different charge generation profiles within the light-harvesting \ce{CuInS2} layer, which strongly affected both the external and internal quantum efficiency. By the reconstruction of these experimental findings with the help of a purely optical model, we were able to draw conclusions on the spatial dependency of the charge collection probability. We provide evidence that only carriers generated within a narrow zone of $\sim$\unit[40]{nm} near the \ce{CuInS2}/ZnO interface contribute to the external photocurrent. The remaining part of the absorber can be considered as ``dead zone'' for charge collection, which reasonably explains the limited device performance and indicates a direction for future research. From the methodical point of view, the optical modeling approach developed in the present work has the advantage that no electrical input parameters are required and is believed to be easily transferable to other material systems.
\end{abstract}

\maketitle

\section{Introduction}
Colloidal nanocrystals~(NCs) have received immense interest as functional materials for photovoltaic applications.\cite{Kramer2011,Graetzel2012,Sargent2012,Borchert2014} Especially, the facile solution processability, the tunability of the band gap due to the quantum-size effect, as well as the ability of multiple exciton generation\cite{Ellingson2005,Semonin2011} make them attractive as absorber material in solar cells. In particular, quantum dots made of lead chalcogenides such as PbS and PbSe have attracted attention in recent years,\cite{Kramer2014,Kim2015} mainly because of their adjustable spectral response, ranging from the visible to the mid-infrared.\cite{Moreels2009} By careful optimization of the optical and electrical properties, remarkable power conversion efficiencies of up to~9.2\% have been achieved for PbS and PbSe colloidal quantum dot solar cells.\cite{Chuang2014,Labelle2014}

However, there are concerns about the potential toxicity and environmental impact of lead compounds, which makes the search for alternative materials an active area of research. One promising class of materials is copper-based nanocompounds such as \ce{Cu2S},\cite{Wu2008} \ce{Cu2ZnSnS4}\cite{Aldakov2013,Guo2009, Guo2010, Suehiro2014, Steinhagen2009} and Cu(In,Ga)(Se,S)$_2$.\cite{Aldakov2013,Guo2008, Stolle2012, Panthani2008, Akhavan2010, Akhavan2010a} These materials have demonstrated reasonable photovoltaic performance when applied to device architectures and fabrication methods known from thin-film solar cells, including the utilization of a cadmium sulfide~(CdS) buffer layer\cite{Wu2008,Guo2009, Guo2010, Suehiro2014, Steinhagen2009, Guo2008, Stolle2012, Panthani2008, Akhavan2010, Akhavan2010a} and extensive post-deposition techniques like high-temperature sintering and selenization.\cite{Guo2009, Guo2010, Guo2008} In contrast, there is still a lack of reports on the application of copper-based NCs to device concepts that are related to the lead chalcogenide quantum dot solar cells, which would enable low-temperature processing and avoid the usage of a CdS buffer layer. Recently, we studied the potential of ternary copper indium disulfide~(\ce{CuInS2}) NCs\cite{Kruszynska2010} as absorber material in such device architectures and successfully implemented them into Schottky solar cells based on a \ce{CuInS2}/metal junction,\cite{Borchert2015} as well as heterojunction devices in combination with n-type zinc oxide~(ZnO) NCs.\cite{Scheunemann2013} However, the device performance remained limited up to now, which we mainly attribute to insufficient charge collection and transport properties.

The collection of charge carriers has been designated as a critical issue in NC solar cells and can become a limiting factor for the external photocurrent. Due to the moderate diffusion length in nanocrystalline solids,\cite{Guglietta2015} the collection of minority carriers via diffusion is rather inefficient.\cite{Kirchartz2015} Because of that, drift collection must account for a large proportion of the generated photocurrent.\cite{Jeong2012,Johnston2008} One promising concept addressing this issue are the so-called depleted heterojunction devices,\cite{Pattantyus2010} in which the commonly p-type light-harvesting NC film is combined with a wide band gap n-type semiconductor such as titania~(\ce{TiO2}) or ZnO. The working principle of these solar cells is based on the formation of a p--n junction, with the depletion region extending over a wide range of the light-harvesting layer, presuming the doping concentrations of both parts of the heterojunction are adequately chosen.\cite{Willis2012} Due to the large extent of the space charge region, where the transport is drift-dominated, the heterojunction devices are clearly advantageous over Schottky junction devices, which posses only a narrow depletion region in the vicinity of the metal contact.\cite{Koleilat2008,Pattantyus2010,Kramer2014} However, the optimum condition of a fully depleted absorber is often not fulfilled, especially when the layer thickness is increased to ensure effective light absorption. Consequently, the charge collection efficiency can become dependent on the spatial position within the light-harvesting layer.\cite{Willis2012, Dibb2013} Substantial ``dead zones'' for charge collection have been reported on quantum dot solar cells,\cite{Barkhouse2010,Willis2012} as well as other low-mobility materials like polymer/fullerene bulk heterojunction films.\cite{Dibb2013} One possibility to overcome these limitations has been proposed by Kramer et al.,\cite{Kramer2011a} who realized quantum dot absorber layers with a band gap gradient to introduce an additional driving force for charge carriers generated beyond the depletion region.

The most commonly used approach to investigate the charge collection properties are electrical characterization techniques in combination with electro-optical modeling.\cite{Kemp2013,Willis2012,Bozyigit2014a,Barkhouse2008} These methods typically require knowledge of various material parameters such as the carrier mobilities, carrier lifetimes, and doping concentrations. However, some of these properties are difficult to determine or simply unknown, in particular when it comes to new materials. The aim of the present work is to utilize an alternative approach to investigate the charge collection deficiencies in \ce{CuInS2}/ZnO heterojunction solar cells. Our approach is based on purely optical simulations, so that only optical properties of the involved materials need to be known. This clearly reduces the amount of necessary input parameters compared to a complete electro-optical description and is believed to make the analysis reliable and easily transferable to other material systems.

In detail, we utilized an inverted device structure\cite{Choi2009,Rath2011,Willis2012} with a p-type substrate and the ZnO NC layer placed between the \ce{CuInS2} and the reflective metal electrode. In that configuration, the functionality of the ZnO is twofold, both as part of the heterojunction and optical spacer layer. It is well-known that the spatial distribution of the optical electric field intensity, which is directly related to the charge carrier generation profile, can be highly structured due to thin-film interference effects.\cite{Pettersson1999,Peumans2003,Burkhard2010,Wilken2015} Because of the optical microcavity formed between the electrodes, systematic variations of the optical spacer thickness can be utilized to manipulate the spatial generation profile within the active layer. The insertion of an optical spacer is a common strategy to improve the photocurrent in organic solar cells.\cite{Kim2006,Gilot2007,Park2009,Dkhil2014} Recently, Kim et al.\cite{Kim2014} reported that, because of the constructive  interference from the optical spacer layer, the inverted architecture can be advantageous for colloidal quantum dot solar cells.

Here, we report on systematic variations of the ZnO thickness to realize different generation profiles within the \ce{CuInS2} layer. These variations had strong impact on the photocurrent and the external quantum efficiency~(EQE), as it is expected for an optical spacer layer. However, we demonstrate that also the internal quantum efficiency~(IQE), which was accurately determined under consideration of optical cavity and parasitic absorption effects,\cite{Burkhard2010,Armin2014} was highly affected by the optical spacer thickness and exhibited a strong dependence on the excitation wavelength. We present an optical model to describe these experimental findings with the help of the spectrally and spatially resolved absorption profiles, provided by simulations based on the transfer-matrix method~(TMM)\cite{Burkhard2010, Pettersson1999, Peumans2003} and making assumptions on the spatial dependence of the collection probability for excess carriers. Using that approach, we present evidence that charges are only collected from a narrow zone in the vicinity of the \ce{CuInS2}/ZnO interface. We interpret this as a hint that the depletion region is not sufficiently extenting into the \ce{CuInS2} layer, which provides a reasonable explanation for the inferior photovoltaic performance.

\section{Results}
\begin{figure}
\includegraphics{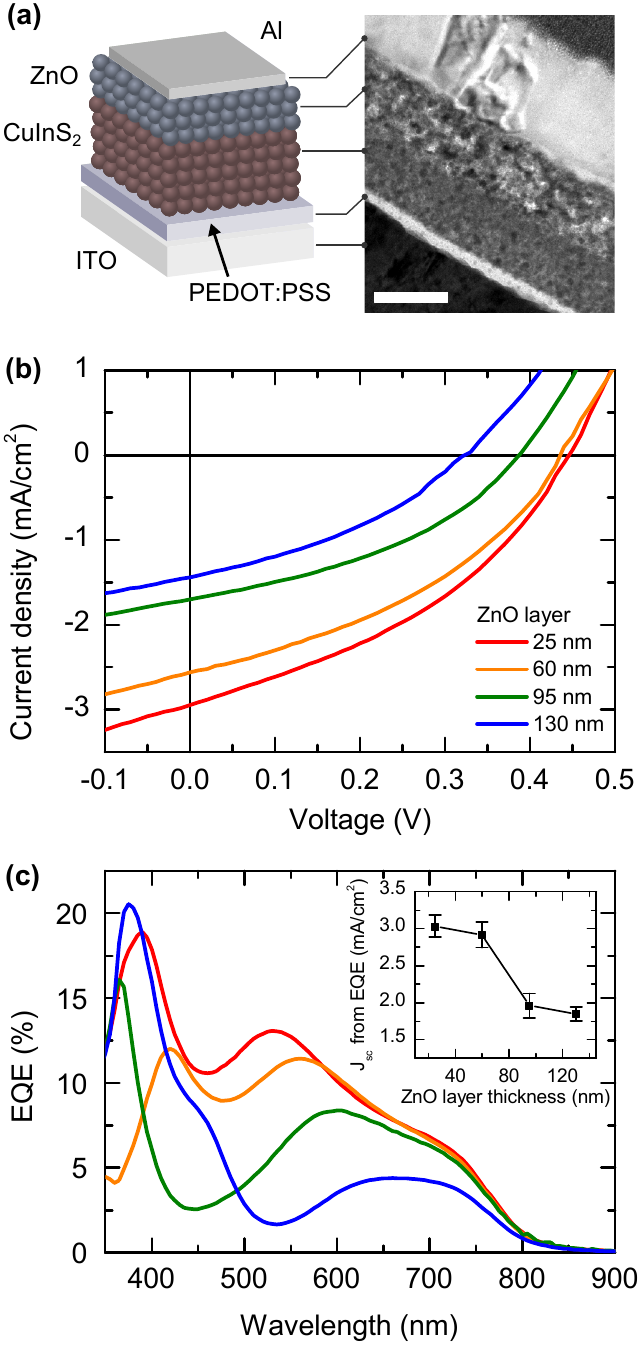}
\caption{(a)~Schematic sketch of the \ce{CuInS2}/ZnO NC solar cells and exemplary cross-sectional TEM micrograph~(scale bar: \unit[100]{nm}). (b)~$J$--$V$ characteristics and (c)~EQE spectra of representative devices for a ZnO layer thickness of \unit[25]{nm}~(red), \unit[60]{nm}~(orange), \unit[95]{nm}~(green), and \unit[130]{nm}~(blue). The inset shows the average photocurrent density~(5-10 devices for each thickness) estimated from the EQE spectra by integration with the AM1.5G standard spectrum.}
\label{fig:Figure1}
\end{figure}

In this work, \ce{CuInS2}/ZnO NC solar cells were fabricated with the structure glass\slash{}indium tin oxide~(ITO)\slash{}poly\-(3,4-ethy\-lene\-di\-oxy\-thio\-phene):poly(sty\-rene\-sulfonate)~(PEDOT:PSS)\slash{}\ce{CuInS2} NCs\slash{}ZnO NCs\slash{}Al. Figure~\ref{fig:Figure1}a shows a schematic sketch of the device architecture, as well as a cross-sectional transmission electron microscopy~(TEM) image of an exemplary device. Both constituents of the heterojunction~(\ce{CuInS2} and ZnO NCs) were synthesized according to previous reports.\cite{Scheunemann2013,Kruszynska2010,Wilken2014,Pacholski2002} Representative TEM micrographs of the as-obtained particles can be found in the Supplemental Material~(Figure~S1). The \ce{CuInS2} NC film, which mainly contributed to the photocurrent generation, was deposited using a one-step spin coating approach without any post-deposition ligand exchange and had a fixed thickness of $\sim$\unit[80]{nm}. In contrast, we varied the thickness of the ZnO layer from 25 to \unit[130]{nm} by adjusting the concentration of the chloroform--methanol solution used for spin coating on top of the \ce{CuInS2}.

Figure~\ref{fig:Figure1}b shows typical current density--voltage~($J$--$V$) characteristics under simulated AM1.5G illumination in dependence on the ZnO layer thickness, and the average photovoltaic characteristics are summarized in Table~\ref{Tab:Table1}. In general, decreasing the thickness of the ZnO layer had a positive effect on the open-circuit voltage~($V_\text{oc}$), while the fill factor~(FF) is only slightly affected and reaches a weak maximum at a medium thickness of \unit[60]{nm}. However, the overall values of FF remain quite low~(0.35--0.41), independent of the ZnO layer thickness, which suggests the presence of severe charge transport limitations associated to the \ce{CuInS2} layer. As a main effect, the thickness of the ZnO layer strongly affects the short-circuit current density~($J_\text{sc}$). Basically, $J_\text{sc}$ is increased by a factor of two by decreasing the thickness of the ZnO layer from 130 to \unit[25]{nm}. In conjunction with the increased $V_\text{oc}$, this results in an improvement of the average power conversion efficiency~(PCE) from 0.2\% to 0.5\%. The champion device obtained within this study reached a PCE of 0.6\%~(see Supplemental Material, Figure~S2), which denotes a threefold improvement compared to our first publication on \ce{CuInS2}/ZnO heterojunction solar cells.\cite{Scheunemann2013} Even though the PCE still remains limited at this stage of research, not least because of the relatively thin photoactive layer of only \unit[80]{nm} used here, it can clearly be seen that the efficiency of the \ce{CuInS2}/ZnO devices is strongly governed by the ZnO layer thickness.

\begin{table}
  \caption{Average Photovoltaic Characteristics of the \ce{CuInS2}/ZnO NC Solar Cells with Varying ZnO Layer Thickness.\protect\footnote{The values reported are mean values, averaged over 10--14 individual devices in each case, and the corresponding standard deviation.}}
  \label{Tab:Table1}
  \centering
  \begin{ruledtabular}
  \begin{tabular}{ccccc}
    ZnO layer & $J_\text{sc}$ & $V_\text{oc}$ & FF & PCE \\
  ~(nm) &~(mA/cm$^2$) &~(V) & &~(\%) \\
   \noalign{\vskip .75mm}
   \hline
   \noalign{\vskip .75mm} 
   25 & 2.9 $\pm$ 0.1 & 0.45 $\pm$ 0.02 & 0.37 $\pm$ 0.02 & 0.48 $\pm$ 0.04\\
   60 & 2.7 $\pm$ 0.2 & 0.46 $\pm$ 0.03 & 0.40 $\pm$ 0.02 & 0.49 $\pm$ 0.06\\
   95 & 1.7 $\pm$ 0.1 & 0.40 $\pm$ 0.04 & 0.38 $\pm$ 0.02 & 0.26 $\pm$ 0.04\\
   130 & 1.4 $\pm$ 0.1 & 0.35 $\pm$ 0.05 & 0.37 $\pm$ 0.03 & 0.19 $\pm$ 0.06\\
   \end{tabular}
   \end{ruledtabular}

\end{table}

In order to investigate the variation of the photocurrent in more detail, we measured the EQE under short-circuit conditions. During the measurements, the samples were continuously illuminated by an additional white light bias source~($\sim$\unit[100]{mW/cm$^2$}) to obtain realistic operation conditions and account for the beneficial effect of light-soaking,\cite{Scheunemann2013} probably related to the ZnO doping density.\cite{Willis2012} Representative EQE spectra for different ZnO layer thicknesses are displayed in Figure~\ref{fig:Figure1}c. For longer wavelengths, the EQE spectra resemble the \ce{CuInS2} absorption spectrum with the onset at about \unit[850]{nm} and the characteristic shoulder at \unit[700]{nm}, whereas, for wavelength below \unit[600]{nm}, strong variations of the absolute intensity, the spectral shape, and the peak positions can be seen in dependence on the ZnO layer thickness. From the EQE spectra, we calculated the photocurrent density to be expected under one sun illumination by integration with the AM1.5G spectrum~(see inset in Figure~\ref{fig:Figure1}c), which coincides well with the trend observed from the $J$--$V$ measurements. Hence, to conclude on the basic photovoltaic characterization, the thickness of the ZnO is found to have strong impact on the photocurrent to be extracted, and the apparent photocurrent variations are mainly correlated with the wavelength-dependent alterations of the EQE spectra in dependence on the ZnO layer thickness.

Basically, there are two possible reasons for the strong wavelength dependence of the extracted photocurrent. First, the spectral absorption within the active \ce{CuInS2} layer and, correspondingly, the generation of excess carriers is modulated by the ZnO layer thickness due to optical interference effects. It is well-known for optically thin solar cell devices with a highly reflective back electrode  that the absorption within the active layer can show strong deviations from Beer--Lambert behavior. This is caused by the constructive or destructive interference of multiple-reflected light components traveling back and forth the device.\cite{Pettersson1999,Peumans2003,Burkhard2010} Thickness variations of any of the non-active layers lead to alterations of the optical electrical field distribution and, hence, are capable to manipulate the active absorption profile.\cite{Wilken2015,Chen2014} Second, also the collection probability for charge carriers could be non-uniformly distributed throughout the active layer. Particularly, in low-mobility materials, the efficiency of charge collection has been found to depend on the position inside the active layer where the charge carriers are generated because of the distance the carriers have to travel, in order to reach the respective electrodes.\cite{Barkhouse2010,Dibb2013,Willis2012}

However, it is noteworthy that the effect of a position-dependent collection probability is usually strongly entangled with the variations of the optical electric field distribution mentioned before. In order to distinguish between both effects, knowledge of the optical electric field distribution $E(\lambda,x)$ within the devices, dependent on the spatial position $x$ and the excitation wavelength $\lambda$, is crucial. Hence, we modeled $E(\lambda,x)$ using one-dimensional TMM simulations.\cite{Burkhard2010,Pettersson1999,Peumans2003} The input parameters of the model include the complex index of refraction $\tilde n = n + i \kappa$, with $n(\lambda)$ and $\kappa(\lambda)$ representing the refractive index and extinction coefficient in dependence on the wavelength, respectively, as well as the thickness of each material layer. We performed detailed analysis of the optical constants~($n$ and $\kappa$) of all materials involved in our devices, using variable angle spectroscopic ellipsometry~(VASE). Details and results of the VASE measurements can be found in the Supplemental Material~(see Figures~S3--S5). On the basis of $E(\lambda,x)$, we calculated device characteristics such as the spectral reflectance and the active absorption within the \ce{CuInS2} layer. From the latter, together with the spectral energy distribution of the illumination~(e.g., the AM1.5G standard spectrum), the charge carrier generation rate and the photocurrent to be expected can be estimated under the assumption of a certain value of the IQE. 

\begin{figure}
\includegraphics{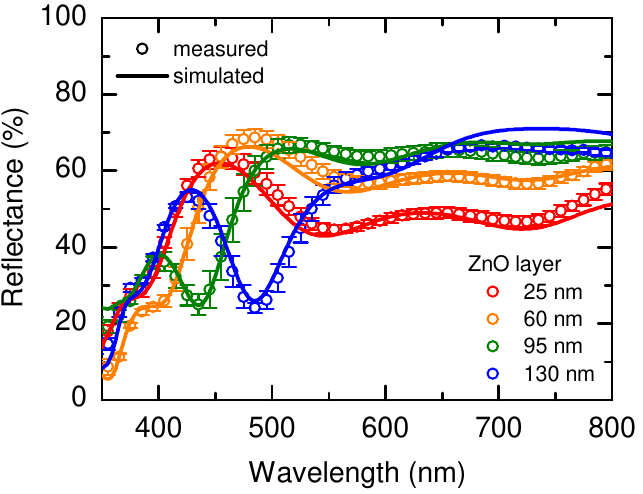}
\caption{Measured~(circles) and simulated~(lines) spectral reflectance for glass\slash{}ITO\slash{}PEDOT:PSS\slash{}\ce{CuInS2}\slash{}ZnO\slash{}Al device stacks with variable thickness of the ZnO layer. The measured data represents an average over several devices and the corresponding standard deviation~(error bars).}
\label{fig:Figure2}
\end{figure}

Figure~\ref{fig:Figure2} shows simulated reflectance spectra of complete solar cell devices, in comparison to experimental data obtained with an integrating sphere. It can clearly be seen that both the spectral shape and the intensity of the reflectance data are reasonably matched by the simulation, which confirms the validity of our TMM model. Only in the case of thicker ZnO layers, slight deviations from the experimental data are visible for wavelengths above \unit[650]{nm}, possibly related to effects that are neglected in the one-dimensional TMM approach such as interface roughness. Nevertheless, the strong variations between the individual reflectance spectra further underline the importance of the ZnO layer thickness on the optical properties of the devices and can be attributed to optical interference effects. 

\begin{figure}
\centering
\includegraphics{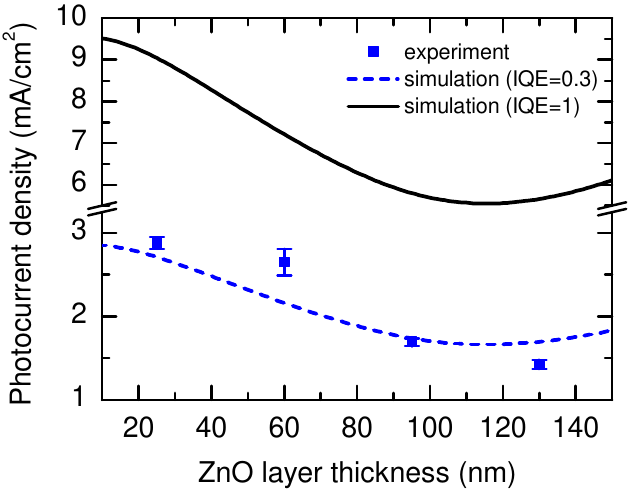}
\hfill 
\caption{Simulated photocurrent density in dependence of the ZnO layer thickness under the assumption of a constant IQE of 100\%~(straight line) and 30\%~(dashed line), respectively, in comparison to the average measured values of $J_\text{sc}$~(symbols).}
\label{fig:Figure3}
\end{figure}

Next, we simulated the photocurrent density to be expected for AM1.5G illumination under the most simple assumption that all absorbed photons would be converted into charge carriers which are successfully extracted at the respective electrodes, corresponding to an IQE of unity. As shown in Figure~\ref{fig:Figure3}~(straight line), the simulation predicts the photocurrent to be strongly affected by the ZnO layer thickness under these premises, with the thinner ZnO layers being more favorable than the thicker ones, which resembles the experimentally observed trend for $J_\text{sc}$~(symbols) and underlines the additional function of the ZnO as optical spacer layer. However, it can be seen that the TMM model clearly overestimates the experimental data if an IQE of unity is assumed. This indicates the presence of severe electrical losses, which probably is the main reason for the limited performance of our devices. As a rough estimate, the experimental data could be described when the IQE is set to a constant, wavelength-independent value of $\sim$30\%~(dashed line), which would imply that $\sim$70\% of the photogenerated carriers do not contribute to the external photocurrent. However, these simple assumptions on the IQE can obviously not explain the experimental data on a quantitative level. Hence, a more detailed knowledge of the apparent IQE is crucial and will be addressed in the following.

In several publications on well-functioning thin film solar cells,\cite{Armin2014,Scharber2013,Armin2013,Law2008} the IQE is found to be independent of the excitation wavelength, provided it is accurately determined with proper consideration of optical cavity effects and parasitic absorption in the non-active layers. However, this assumption is not valid in general, as has been discussed by Burkhard et al.\cite{Burkhard2009} The authors attributed the reported spectral dependence of the IQE to incomplete exciton harvesting, which demonstrates that IQE data can give valuable insight into the device physics. Therefore, we determined the spectral IQE of our \ce{CuInS2}/ZnO solar cells in dependence of the ZnO layer thickness. At this point, we would like to point out that the accurate determination of the IQE is a crucial issue,\cite{Burkhard2010,Armin2014} and errors during the evaluation process can lead to substantial misinterpretation of experimental results.\cite{Armin2013,Scharber2013} In general, the IQE is defined as the ratio of the number of charge carriers collected at the electrodes to the number of photons absorbed throughout the active layer. Because of the fact that the latter cannot be measured directly, several techniques are known for the evaluation of the IQE in thin film solar cells. The most simple approach is to calculate the ratio between the measured and simulated photocurrent, neglecting any possible spectral dependencies.\cite{Chang2013,Jasieniak2011} To obtain spectrally resolved data, the IQE is commonly calculated from EQE spectra that are normalized to the absorption in the active layer, either simulated by TMM-based optical modeling\cite{Semonin2011,Law2008} or estimated by transmittance and/or reflectance measurements. However, it has recently been shown in a comprehensive synopsis by Armin et al.\cite{Armin2014} that none of those methods is able to properly consider (i)~the optical cavity effects caused by thin-film interference, (ii)~the parasitic absorption in contact and interfacial layers, as well as (iii)~scattering and surface roughness altogether. In consequence, we determined the IQE following the procedure suggested in Ref.~\citenum{Armin2014}, assuming the transmission through the complete solar cells to be negligible~(which is fulfilled because of the \unit[120]{nm} thick Al electrode). For this purpose, we used the spectral reflectance $R(\lambda)$ of the devices~(see Figure~\ref{fig:Figure2}), which was measured using an integrating sphere to include the diffuse reflectance caused by scattering effects, as well as the parasitic absorption $\text{PA}(\lambda)$ within the non-active layers, which was simulated using the TMM approach described above. Finally, we calculated the spectrally dependent IQE from the measured EQE spectra by

\begin{equation}
\text{IQE}(\lambda)= \frac{\text{EQE}(\lambda)}{1-R(\lambda)-\text{PA}(\lambda)}.
\label{eq:eq1}
\end{equation}

For simplicity reasons, light absorption in the optical spacer layer was considered as \textit{parasitic} absorption within these calculations, although there is a possibility for carriers generated in the ZnO to become separated at the hetojunction and contribute to the photocurrent. However, due to the wide band gap of ZnO, this process is only of relevance for wavelengths below \unit[360]{nm}. Correspondingly, we found that the contribution of the ZnO was only 1-2\% of the total photocurrent.

\begin{figure}
\centering
\includegraphics{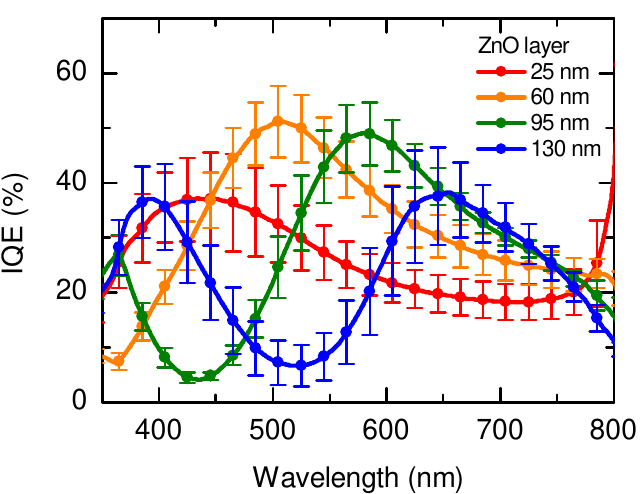}
\hfill
\caption{Spectrally resolved IQE of \ce{CuInS2}/ZnO heterojunction solar cells with ZnO layer thicknesses ranging from 25 to \unit[130]{nm}, averaged over 7--10 individual devices in each case. The error bars represent the corresponding standard deviation.}
\label{fig:Figure4}
\end{figure}

Figure~\ref{fig:Figure4} depicts the results according to Eq.~(\ref{eq:eq1}) in dependence of the ZnO layer thickness. Contrary to another publication on PbSe quantum dot solar cells,\cite{Law2008} where almost flat IQE spectra are reported over a large range of wavelengths, a strongly oscillating behavior can be seen for our \ce{CuInS2}/ZnO heterojunction devices, with the minimum and maximum positions shifting toward longer wavelengths with increasing ZnO layer thickness. In particular, pronounced minima of the IQE are visible at wavelengths of $\sim$\unit[360]{nm}, $\sim$\unit[435]{nm}, and $\sim$\unit[520]{nm} for the \unit[60]{nm}, \unit[95]{nm}, and \unit[130]{nm} thick ZnO layers, respectively, where the IQE drops down to almost zero. Since the IQE is defined as the ratio of collected charges to the number of photons that are actually absorbed within the active layer, which is supposed to be equal to the number of charge carriers generated, the IQE itself is not influenced by optical effects, but rather related to the efficiency of the elementary processes of (i)~charge transport toward the heterojunction, (ii)~charge separation within the heterojunction, (iii)~charge transport toward the respective electrodes, and~(iv)~charge extraction at the electrodes, which will be summarized by the term \textit{charge collection} in the following. In view of this, the question arises, why the charge collection should exhibit such a strong dependence on the excitation wavelength.

To answer this question, one has to take into account that, for the reasons outlined above, the spectral absorption profile is highly non-uniformly distributed throughout the active layer, in a way that photons with different energies, corresponding to different excitation wavelengths, are absorbed at different spatial positions. Hence, it is reasonable to take a closer look at the spectrally and spatially resolved photon absorption profile $A(\lambda,x)$ at first, which is directly proportional to the~(simulated) distribution of $E(\lambda,x)$, but weighted with the optical constants of the active layer:\cite{Pettersson1999, Braunstein1976}

\begin{equation}
A(\lambda,x) = \frac{4\,\pi}{\lambda}\,n(\lambda)\,\kappa(\lambda)\,\vert E(\lambda,x) \vert^2 \text{.}
\label{eq:eq2}
\end{equation}

\begin{figure*}
\includegraphics{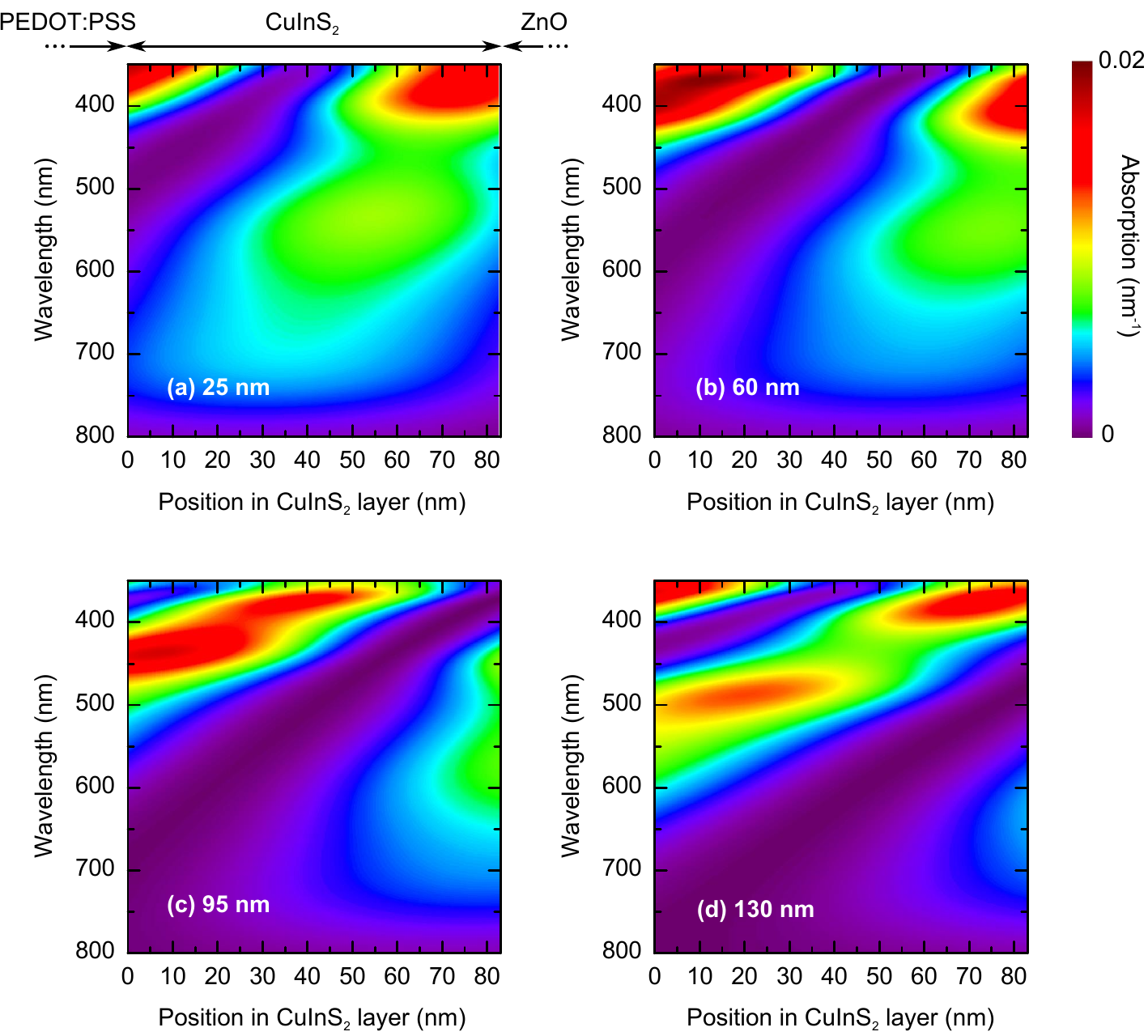}
\caption{Simulated photon absorption profiles $A(\lambda,x)$ within the \ce{CuInS2} layer for different ZnO layer thicknesses of (a)~\unit[25]{nm}, (b)~\unit[60]{nm}, (c)~\unit[95]{nm}, and (d)~\unit[130]{nm}. The position of \unit[0]{nm}~(\unit[82]{nm}) corresponds to the PEDOT:PSS~(ZnO) interface.}
\label{fig:Figure5}
\end{figure*}

Note that $E(\lambda,x)$ is normalized to the incident electric field, so that Eq.~(\ref{eq:eq2}) represents the absorption per unit length. Figure~\ref{fig:Figure5} displays the resulting $A(\lambda,x)$ profiles for varying thickness of the ZnO layer. As a first impression from these plots, the overall absorption appears higher for thin ZnO layers, which is consistent with the decrease of the photocurrent with increasing ZnO thickness, as shown before in the results of the $J$--$V$ and EQE measurements. However, much more interesting is the comparison of $A(\lambda,x)$ with the spectral dependence of the IQE~(see Figure~\ref{fig:Figure4}). It can be seen that a minimum of the IQE is observed for wavelengths at which the fraction of absorbed light is high in the vicinity of the hole-collecting electrode~(PEDOT:PSS), whereas, high IQE values correspond with high absorption in the vicinity of the \ce{CuInS2}/ZnO interface. For instance, if considering a ZnO layer thickness of \unit[95]{nm}~(Figure~\ref{fig:Figure5}c), a large fraction of light is absorbed nearby the PEDOT:PSS interface at an excitation wavelength of $\sim$\unit[430]{nm}, which directly corresponds to the minimum of the IQE in that wavelength region. In contrast, at a wavelength of $\sim$\unit[580]{nm}, absorption mainly takes place close the heterojunction, corresponding to the maximum of the IQE spectrum. Similar trends can be observed for all other ZnO layer thicknesses. All in all, these direct correlations between IQE($\lambda$) and $A(\lambda,x)$ clearly suggest that the probability for excess carriers to contribute to the external photocurrent is closely interconnected with the spatial position where they are originally generated within the active layer. In other words, the spectral dependence of the IQE directly translates into a spatially dependent charge collection probability, $\eta_\text{cc}$ = $\eta_\text{cc}(x)$. The collection probability can be interpreted as spatial weighting function for the photon absorption profiles to yield the spectral dependence of the EQE, which then can be written as the product of $A(\lambda,x)$ and $\eta_\text{cc}(x)$, integrated over the thickness $d$ of the active layer:

\begin{equation}
\text{EQE}(\lambda)=\int_0^d\,A(x,\lambda) \, \eta_\text{cc}(x)\,\text{d}x.
\label{eq:eq3}
\end{equation}

\begin{figure*}
\includegraphics{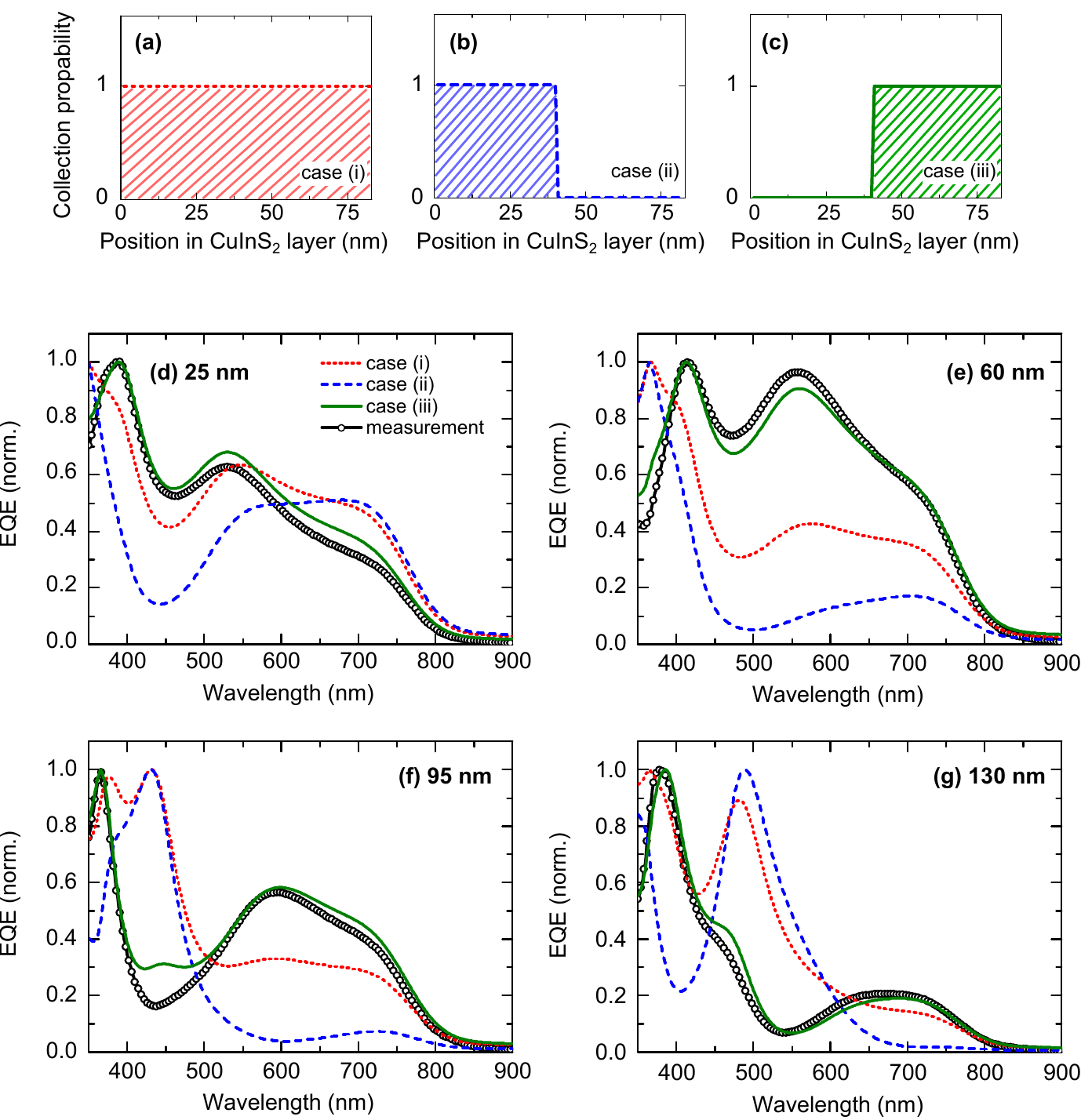}
\caption{(a)--(c)~Spatially dependent charge collection probability functions $\eta_\text{cc}(x)$ that were assumed for the reconstruction of the measured EQE spectra. The position of \unit[0]{nm}~(\unit[82]{nm}) corresponds to the PEDOT:PSS~(ZnO) interface. (d)--(g)~Reconstructed EQE spectra corresponding to the cases displayed in panel~a~(dotted lines), panel~b~(dashed lines), and panel~c~(straight lines), in comparison to the experimental data~(symbols) for ZnO layer thicknesses of (d)~\unit[25]{nm}, (e)~\unit[60]{nm}, (f)~\unit[95]{nm}, and (g)~\unit[130]{nm}.}
\label{fig:Figure6}
\end{figure*}

In principle, $\eta_\text{cc}(x)$ could now be calculated from the measured EQE spectra and the simulated absorption profiles by direct inversion of Eq.~(\ref{eq:eq3}). However, this problem is normally overdetermined, and the common numerical solution algorithms tend to convergence to unphysical solutions. Therefore, we used a straightforward strategy instead and assumed three simple cases for the spatial dependence of $\eta_\text{cc}$ that are illustrated in Figure~\ref{fig:Figure6}a--c: (i)~constant collection throughout the whole active layer, (ii)~constant collection only in the half of the active layer adjacent to the PEDOT:PSS interface, and (iii)~constant collection only in the half of the active layer adjacent to the ZnO interface. For the latter two cases, we approximated the charge collection probability as step functions that are centered in the middle of the active layer, ranging from 0 to 1.

From the model functions for $\eta_\text{cc}(x)$ and the simulated photon absorption profiles, we calculated the EQE spectra according to Eq.~(\ref{eq:eq3}). Figure~\ref{fig:Figure6}d--g shows the results of these calculations in comparison to the measured EQE spectra for the different ZnO layer thicknesses under investigation. The data is presented on a normalized scale, because, at first, we were only interested in the reconstruction of the spectral shape. Obviously, case~(i) and~(ii) are not able to describe the situation properly and large deviations can be seen between the measured and modeled EQE spectra. In contrast, both the shape and the relative intensity between the spectral features of the measured EQE are well reproduced in case~(iii). This finding is a clear indication that only a limited region of the active layer~(``collection zone'') near the heterojunction contribute to the photocurrent, while the remaining part of the active layer can be considered as ``dead zone'' for charge collection.

\begin{figure}
\includegraphics{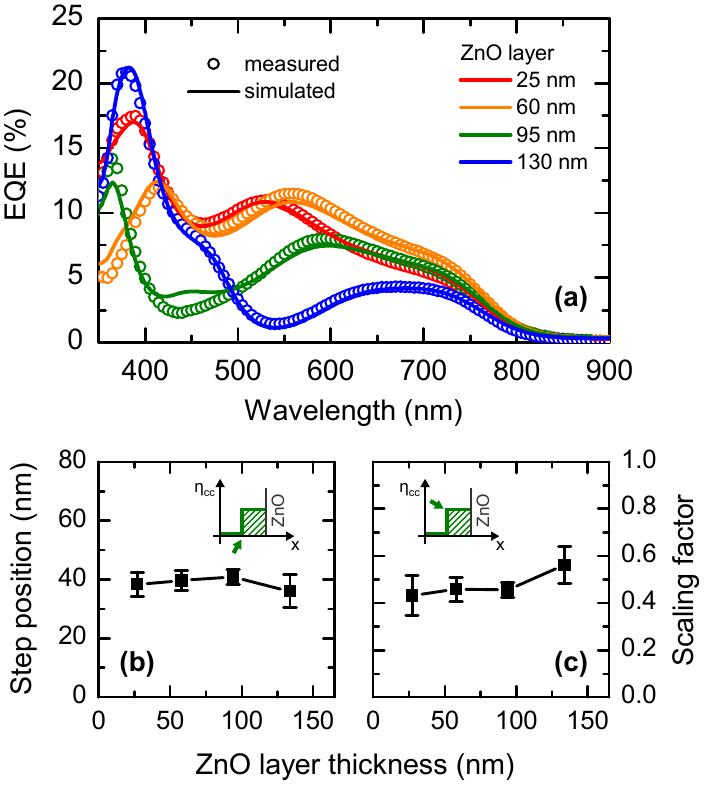}
\caption{Results for the optimization of the step-like charge collection function $\eta_\text{cc}(x)$ in dependence on the ZnO layer thickness. (a)~Comparison of the measured and reconstructed EQE spectra for representative devices on an absolute scale. (b)~Position of the step edge in terms of the distance from the \ce{CuInS2}/ZnO interface. (c)~Scaling factor related to the height of the step function. The data presented in panels~b and c is an average of several devices with the corresponding standard deviation~(error bars).}
\label{fig:Figure7}
\end{figure}

For a more detailed investigation of the apparent collection and dead zones, we now concentrate on case~(iii) and adjusted the step-function such that the measured EQE spectra are also described on a quantitative manner. Therefore, we used a two-step approach: first, we optimized the spatial position of the step edge by maximizing the linear correlation coefficient between the measured and modeled EQE spectra; second, we adjusted the height of the step function by introducing a scaling factor, which was optimized using a least mean square algorithm. Finally, we obtained an excellent agreement between the modeled and measured EQE spectra on an absolute scale, as it is demonstrated in Figure~\ref{fig:Figure7}a. The optimization procedure was performed individually for several devices with different ZnO layer thickness, and the average results are displayed in Figure~\ref{fig:Figure7}b,c. It can be seen from the optimization of the step position~(Figure~\ref{fig:Figure7}b), which is directly correlated with the spectral shape of the EQE, that the collection zone has an average width of $\sim$\unit[40]{nm}, independent of the thickness of the ZnO layer. Instead, for the scaling factor~(Figure~\ref{fig:Figure7}c), which affects only the height of the EQE spectra, a slight increase can be seen with increasing ZnO layer thickness, however, due to the relatively large error bars, our data provide no statistical evidence for this tendency. Nevertheless, the scaling factors remain significantly smaller than 1, which indicates that even within the collection zone only a part of the generated carriers contributes to the external photocurrent.

\begin{figure}
\centering
\includegraphics{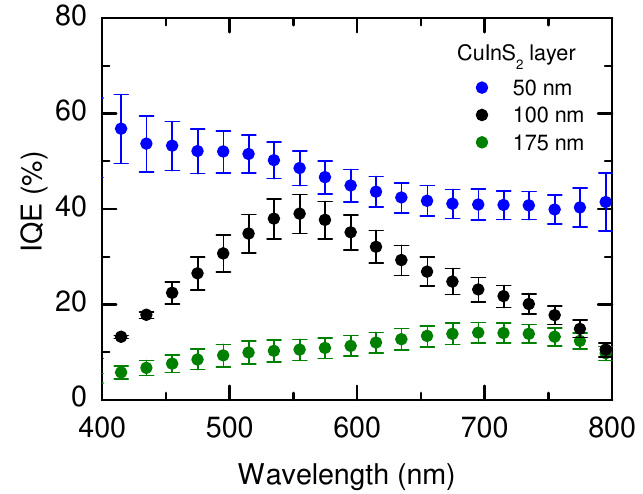}
\hfill
\caption{Spectrally resolved IQE of \ce{CuInS2}/ZnO heterojunction solar cells with variable \ce{CuInS2} thickness ranging from 50 to \unit[175]{nm} and a fixed ZnO thickness of \unit[70]{nm}. The data represents an average of 4--6 individual devices in each case and the corresponding standard deviation~(error bars).}
\label{fig:Figure8}
\end{figure}

To further confirm our analysis, we also performed a variation of the \ce{CuInS2} layer thickness while the ZnO layer was kept at a constant thickness of \unit[70]{nm}. Figure~\ref{fig:Figure8} depicts the IQE for 50, 100, and \unit[175]{nm} thick \ce{CuInS2} layers. For the thin \ce{CuInS2} layers~(\unit[50]{nm}), the IQE appears fairly flat over the range of wavelengths considered, being an indication that the charge collection is only weakly dependent on the spatial position. This corresponds well to the fact that the collection zone is nearly extending over the whole absorber in that case. At a medium \ce{CuInS2} thickness~(\unit[100]{nm}), the IQE is highly structured as shown before~(compare Figure~\ref{fig:Figure4}), whilst for further increase of the \ce{CuInS2} thickness~(\unit[175]{nm}), charge collection becomes extremely inefficient with a slight increase of the IQE for larger wavelengths. This can be expected for thick absorber layers because short wavelength photons~(entering from the ITO/PEDOT:PSS side) are not able to penetrate deep into the device. Also the total decrease of the IQE appears reasonable in the light of our previous findings because the thicker the absorber gets, the larger is the proportion of the dead zone for charge collection on the total absorber thickness.

Using our optical simulation approach, we also reconstructed the EQE spectra for variable \ce{CuInS2} layer thickness and the results can be found in the Supplemental Material~(Figures~S6 and S7). It can be seen that the width of the collection zone is nearly independent on the thickness of the \ce{CuInS2} layer, which is in agreement with the findings outlined above. Instead of that, the optimization procedure resulted in a steady decrease of the scaling factor with increasing \ce{CuInS2} thickness, ranging from 0.6 to 0.4. This appears reasonable when considering the longer distance the separated holes have to travel to reach the PEDOT:PSS interface, resulting in an increased probability for recombination prior to the extraction.

\section{Discussion}
To summarize our results up to this point, we were able to identify substantial dead zones for charge collection as significant loss mechanism in our \ce{CuInS2}/ZnO heterojunction solar cells. Moreover, even within the collection zone of the active layer, the probability for excess carriers to contribute to the external current is limited to $\sim$50\%. In turn, the combination of both effects can provide a reasonable explanation of why the observed IQE values are significantly lower than 100\% for all wavelengths~(see Figure~\ref{fig:Figure4}). In the following, we will discuss the possible origin for these deficiencies in charge collection. Dead zones in NC solar cells have already been reported by Barkhouse et al.\cite{Barkhouse2010} However, a different device architecture was utilized in that study, with the ITO electrode applicated on top of the NC layer by sputtering, and the reduced charge collection was attributed to recombination centers induced by sputter damage during the electrode deposition process, which can be excluded in our case. 

Another explanation would be that the charge collection probability corresponds to the spatial extension of the depletion region within the \ce{CuInS2}/ZnO devices. If the depletion region width $w$ becomes smaller than the thickness of the active layer~($w < d$), there remains a field-free, quasi-neutral region where the charge transport is controlled by diffusion.\cite{Koleilat2008} Typically, charge collection is found to be less efficient in a diffusion-limited region than within the depletion region, where the transport is drift-dominated. The transport properties in these regimes are characterized by the diffusion length $L_\text{diff}$ and the drift length $L_\text{drift}$, respectively, with both quantities being directly dependent on the mobility--lifetime~($\mu\tau$) product.\cite{Hegedus2004,Jeong2012} Very recently, Kirchartz et al.\cite{Kirchartz2015} proposed design rules for the optimal ratio $w/d$ as a function of the magnitude of $\mu\tau$. For relatively low values of the $\mu\tau$ product, supposed to be the relevant case here, full depletion of the active layer~($w/d \sim 1$) is clearly desirable because diffusion-limited collection is highly ineffective in that case with quite low values of $L_\text{diff}$ in the range of a few nm and $L_\text{drift} > L_\text{diff}$. In that context, Dibb et al.\cite{Dibb2013} investigated the charge collection in organic bulk heterojunction solar cells with a doped active layer and $w/d < 1$. The authors found that the collection efficiency in the diffusion-limited region became virtually zero and, similar to our approach, proposed a step-like charge collection function with 100\% collection within the depletion region and 0\% collection in the remaining quasi-neutral region to model the external photocurrent.\cite{Dibb2013} 

If we now come back to our bilayer \ce{CuInS2}/ZnO NC solar cells, the scenario that the devices are not fully depleted appears reasonable because of the fact that excess carriers are only collected in a relatively narrow region in the vicinity of the heterojunction. Accordingly, the dead zone would correspond to a quasi-neutral region through which minority carriers have to diffuse. The quantities $\mu\tau$ and $L_\text{diff}$ are supposed to be particularly small within the \ce{CuInS2} layer due to the not yet optimized ligand treatment, a fact that is further supported by the relatively low fill factors and high values of the series resistance observed for our devices. Hence, recombination is quite likely in the quasi-neutral zone, with the result that none of the excess carriers generated in that region are able to reach the heterojunction and contribute to the photocurrent. However, we would like to emphasize that it is difficult to identify the width of the collection zone directly with the width of the depletion region in the \ce{CuInS2} layer, because the rectangular profiles applied here are certainly a simplification of the real collection probability functions.

Limitations of the device performance related to the extent of the depletion region into the light-harvesting layer have already been reported on heterojunction NC solar cells.\cite{Willis2012,Rath2011} In general, the magnitude of $w$ is related to the doping concentrations of the p-type and n-type material, with the depletion region being more extended in the less doped layer. In case of PbS/ZnO NC devices, Willis et al.\cite{Willis2012} showed that the portion of the depletion region lying within the p-type PbS film is increased by photodoping of the ZnO layer with UV light. The latter is caused by the photoinduced release of electrons, trapped in a surface system such as chemisorbed oxygen,\cite{Wilken2014} with the result of an increased ZnO doping density, which is near or greater than that of the PbS.\cite{Willis2012} Regarding our \ce{CuInS2}/ZnO solar cells, the exposure to UV-containing illumination~(light-soaking) was found to have positive effect on the device performance as well,\cite{Scheunemann2013} which is likely to be related to photodoping of the ZnO layer. However, even though light soaking was thoroughly conducted prior to the electrical measurements in the present study, the ZnO doping concentration probably was not high enough to provide full depletion of the \ce{CuInS2}.

There remains the question of why the collection probability for carriers originating from the collection zone is restricted to $\sim$50\%. In principle, only the drift-dominated transport is influenced by an external electric field.\cite{Hegedus2004,Kemp2013} In order to investigate the electric field dependence of the charge collection in more detail, we measured EQE spectra with an additionally applied bias voltage, and the results are depicted in Figure~S8 in the Supplemental Material. It can be seen that the EQE is improved under reverse bias voltage, indicating that more photogenerated carriers are collected. Most notably, only the height of the spectra is affected, whereas, the shape appears more or less the same. According to our previous results, this would imply that the bias voltage has a strong effect on the efficiency of charge collection within the collection zone, but only slightly affects its extension throughout the active layer. In the picture that the devices are not fully depleted, the improved charge collection could be attributed to a voltage-dependent enlargement of $L_\text{drift}$ within the depletion region, with the result that more carriers are swept outside the device. However, one would also expect a variation of $w$ with the external voltage,\cite{Dibb2013,Hegedus2004,Kemp2013} which is not clearly seen in our data. Despite that, it is worth noting that even under a bias voltage of \unit[-1]{V}, where the improvement of the EQE saturate, it is still less than one would expect for 100\% collection within the collection zone. Hence, there are obviously other electrical loss mechanism present that are not considered here, for instance, energetic barriers for charge extraction/injection at one of the electrodes,\cite{Brown2011} or trap states related to the NC films.\cite{Kemp2013,Bozyigit2014b}

\begin{figure}
\centering
\includegraphics{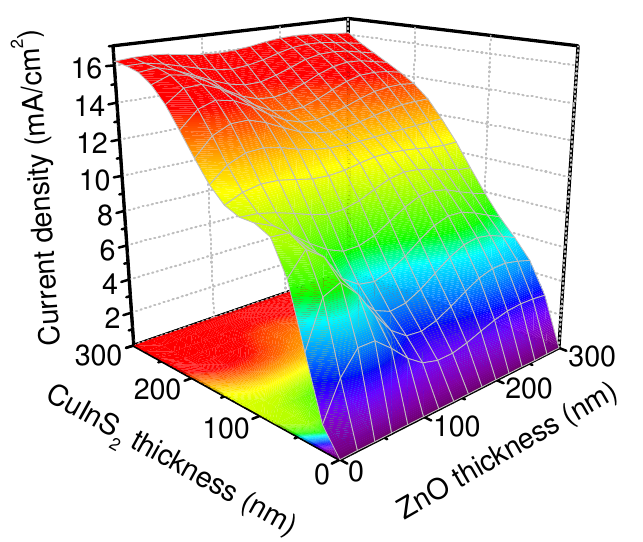}
\caption{Simulated photocurrent density in dependence on the \ce{CuInS2} and ZnO layer thickness for AM1.5G spectral irradiance and under the assumption of ideal charge collection properties~($\text{IQE} = 100\%$).}
\label{fig:Figure9}
\end{figure}

Finally, we would like to point out that our results will need to be corroborated by means of independent electrical characterization techniques. However, it is evident from the present study that charge collection is an important bottleneck in our \ce{CuInS2}/ZnO heterojunction solar cells and has to be improved to compete with more established systems like lead chalcogenide quantum dots, especially when the thickness of the light-harvesting \ce{CuInS2} layer is increased, which is necessary to enhance the total photon absorption. Provided that such improvement of the collection efficiency is possible, high photocurrents are to be expected, as it can be seen from Figure~\ref{fig:Figure9}, where the simulated upper limit of the photocurrent~($\text{IQE} = 100\%$) is shown as a function of the \ce{CuInS2} and ZnO layer thickness. In view of this, the \ce{CuInS2}/ZnO system has the potential to reach reasonable device performance.

\section{Conclusions}
In summary, we have investigated Cd- and Pb-free \ce{CuInS2}/ZnO NC solar cells with the structure ITO\slash{}PEDOT:PSS\slash{}\ce{CuInS2} NCs\slash{}ZnO NCs\slash{}Al. The ZnO layer was found to have a double function, both as constituent of the heterojunction and as optical spacer. Thus, varying the ZnO thickness enabled us to modulate the excess carrier generation profiles, which strongly affected the photocurrent, as well as spectral shape of the EQE. To understand these variations in more detail, we investigated the IQE under careful consideration of optical cavity effects and the parasitic absorption within the non-active layers. Most notably, the IQE was found to exhibit a pronounced dependence on the excitation wavelength, which could be translated into a spatial dependence of the charge collection probability.

With the help of modeled photon absorption profiles provided by TMM simulations, and making simple assumptions on the charge collection probability as a function of the position within the active layer, we reconstructed the measured spectral dependencies of the EQE. Using this optical approach, which does not require additional assumptions on electrical properties of the involved materials, we were able to identify pronounced dead zones for charge collection in the vicinity of the PEDOT:PSS interface. We interpret this as a hint that our \ce{CuInS2}/ZnO devices are not fully depleted, with the result that excess carriers generated beyond the depletion region are not able to contribute to the external photocurrent because of the short minority carrier diffusion length. Moreover, even within the collection zone, the collection probability was found to be restricted to $\sim$50\% under short-circuit conditions. 

Altogether, these deficiencies in charge collection provide a reasonable explanation for the limited device performance up to now and highlight the starting points for future improvements. Finally, from a more general point of view, we would like to emphasize that the methods presented here to study the spatial dependence of charge collection are expected to be easily transferable to other absorber systems and kind of photovoltaic devices.

\section*{Experimental Section}
\subsection*{Nanoparticle Synthesis}
Pyramidally shaped wurtzite-type \ce{CuInS2} NCs with an edge length ranging from 12 to \unit[18]{nm} were synthesized using a hot-injection technique, as described previously.\cite{Scheunemann2013,Kruszynska2010} Briefly, a 1:1~(v/v) mixture of 1-dodecanethiol and \textit{tert}-dodecanethiol was rapidly injected into a reaction vessel containing copper acetate, indium(III) acetate, trioctylphosphine oxide, and oleylamine at \unit[220]{$^\circ$C} under argon atmosphere. The mixture was then heated up to \unit[250]{$^\circ$C}, and the reaction was stopped after \unit[1]{h} by cooling to room temperature. Subsequently, the as-obtained particles were precipitated and washed with ethanol to remove residuals from the synthesis. To exchange the long-chained ligands used in the synthesis, the NCs were redissolved in 1-hexanethiol and heated under permanent stirring for \unit[24]{h} at \unit[100]{$^\circ$C}. Afterward, the particles were precipitated with ethanol and re-dispersed in chlorobenzene. Quasi-spherically shaped ZnO NCs with average diameter of $\sim$\unit[5]{nm} were synthesized from zinc acetate dihydrate and potassium hydroxide, according to previous reports.\cite{Wilken2014,Pacholski2002} The as-obtained particles were filtered through a \unit[0.2]{$\mu$m} syringe filter and dissolved in a 9:1~(v:v) mixture of chloroform and methanol.

\subsection*{Solar Cell Fabrication}
\ce{CuInS2}/ZnO solar cells were fabricated on cleaned and patterned ITO-coated glass substrates~(Pr\"azisions Glas \& Optik GmbH, Germany; sheet resistance $\leq$\unit[10]{$\Omega$/sq}) covered with PEDOT:PSS~(Clevios P VP AI 4083), as described previously.\cite{Scheunemann2013} The \ce{CuInS2} layer was deposited in a nitrogen filled glove box by spin coating a \unit[60]{mg/ml} dispersion at \unit[600]{min$^{-1}$}, followed by a drying step at \unit[3000]{min$^{-1}$}. Subsequently, the ZnO layer was spin-coated at \unit[1500]{min$^{-1}$}. Different ZnO layer thicknesses of $\sim$25, $\sim$60, $\sim$95, and $\sim$\unit[130]{nm} were obtained by using dispersions with nominal concentrations of 7.5, 15, 22.5, and \unit[30]{mg/ml}. Finally, the \unit[120]{nm} thick Al electrode was thermally evaporated under high vacuum~(\unit[$10^{-6}$]{mbar}). For the thickness variation of the \ce{CuInS2} layer, the concentration of the \ce{CuInS2} NC dispersion was varied between 17.5 and \unit[120]{mg/ml}. The samples were annealed at \unit[180]{$^\circ$C} both after deposition of the \ce{CuInS2} and the ZnO layer. The active area~($\sim$\unit[14]{mm$^2$}) was delimited by the geometric overlap of the ITO and metal electrode and precisely measured for each individual device using a stereoscopic microscope equipped with a digital camera~(Cascade Microtech EPS150COAX).

\subsection*{Characterization Methods}
For the cross-sectional TEM image, a lamella was prepared with a focused ion beam system~(FEI Helios NanoLab 600i) and inspected by high-resolution TEM~(Jeol JEM-2100F). Film thicknesses were determined using a stylus profiler~(Veeco Dektak 6M). $J$--$V$ characteristics were recorded under ambient conditions with a semiconductor characterization system~(Keithley 4200). The samples were illuminated using a class AAA simulator~(Photo Emission Tech.), providing a simulated AM1.5G spectrum. The light intensity was adjusted to \unit[100]{mW/cm$^2$} using a calibrated silicon solar cell. The spectral mismatch factor was 1.037~(not taken into account). Prior to the $J$--$V$ measurements, the samples were subjected to continuous illumination for $\sim$\unit[10]{min}~(light-soaking). EQE measurements were performed using a custom-built setup~(Bentham PVE300), equipped with a dual xenon/quartz halogen lamp and a Czerny--Turner monochromator~(Bentham TMc300) providing the monochromatic probe light. The light source was modulated at a frequency of \unit[230]{Hz} by an optical chopper, and the photocurrent was recorded under short-circuit conditions with a lock-in amplifier~(Stanford Research Systems SR830). During the EQE measurements, the samples were continuously biased by an additional white light xenon lamp with an intensity adjusted to $\sim$\unit[100]{mW/cm$^2$}. Using this white light bias source, the samples were light-soaked prior to the measurements until saturation of the photocurrent was reached. The EQE setup, additionally equipped with a \unit[150]{mm} integrating sphere~(Bentham DTR6), was also used to measure the spectral reflectance of complete device stacks at an incidence angle of 8$^\circ$. A light trap was used to distinguish between specular and diffuse reflectance. Because our device showed only little diffuse reflection~($\leq 3\%$), we used a highly specular reflecting silicon wafer with known optical constants as calibration standard for the reflectance measurements. Transmittance through the devices was found to be negligible at all wavelengths.

\subsection*{Determination of the Optical Constants}
VASE measurements were carried out in ambient air on a rotating analyzer ellipsometer with auto retarder~(J.~A.~Woollam) under at least 3 angles of incidence and in a wavelength range between 280 and \unit[1700]{nm}~(\unit[5]{nm} step size). Either silicon wafers or float glass with known optical properties were used as substrate. The optical constants~(i.e., $n(\lambda)$ and $\kappa(\lambda)$) were determined by means of modeling the ellipsometric data by Kramers--Kronig consistent dispersion models using the supplied software WVASE32~(J. A. Woollam). Details regarding the measurements and data modeling can be found in the Supplemental Material. In case of aluminum, the optical constants could not be determined by ellipsometry because of the oxide layer formation on the surface in ambient air. Therefore, we used literature values from Ref.~\citenum{Rakic1998} instead, which were validated with the spectral reflectance of glass/Al samples measured from the protected glass side~(see Supplemental Material, Figure~S5).

\subsection*{Optical Simulations}
The optical electric field distribution $E(\lambda,x)$ within the devices~(normalized to the incoming electric field) was simulated by the help of an one-dimensional TMM model, based on a MATLAB algorithm developed by Burkhard and Hoke.\cite{Burkhard2010} Therefore, each of the material layers was parameterized by its thickness and optical constants~($n$ and $\kappa$). On the basis of $E(\lambda,x)$, other device characteristics such as the spectral reflectance $R(\lambda)$, the photon absorption profile $A(\lambda,x)$ within the active layer, as well as the parasitic absorption $\text{PA}(\lambda)$ within the non-active layers was calculated. Photocurrent densities were estimated by $j_\text{ph} = -q \int_0^d G(x)\,\text{d}x$, where $q$, $d$, and $G(x)$ represent the elementary charge, the active layer thickness, and the charge generation rate, respectively. $G(x)$ was calculated from $A(\lambda,x)$ and the AM1.5G spectral irradiance distribution.

\begin{acknowledgements}
The authors thank Janet Neerken for practical support, Michael Richter for assistance with preliminary work on the optical simulations, and Jan Keller for help with the TEM lamella preparation. Financial support from the EWE-Nachwuchsgruppe ``D\"unnschichtphotovoltaik'' by the EWE AG, Oldenburg, Germany, is gratefully acknowledged.
\end{acknowledgements}

\section*{Author Contributions}
D.S. and S.W. contributed equally to this work.

\bibliography{references}

\end{document}


\title{Supplemental Material for ``Investigation of the Spatially Dependent Charge Collection Probability in \ce{CuInS2}/ZnO Colloidal Nanocrystal Solar Cells''}

\author{Dorothea Scheunemann}
\email{dorothea.scheunemann@uol.de}
\author{Sebastian Wilken}
\email{sebastian.wilken@uol.de}
\author{J\"urgen Parisi}
\author{Holger Borchert}

\affiliation{Institute of Physics, Energy and Semiconductor Research Laboratory, Carl von Ossietzky University of Oldenburg, 26111 Oldenburg, Germany}

\maketitle
\tableofcontents

\section{TEM Images}
\begin{figure}[h!]
\includegraphics[width=0.75\textwidth]{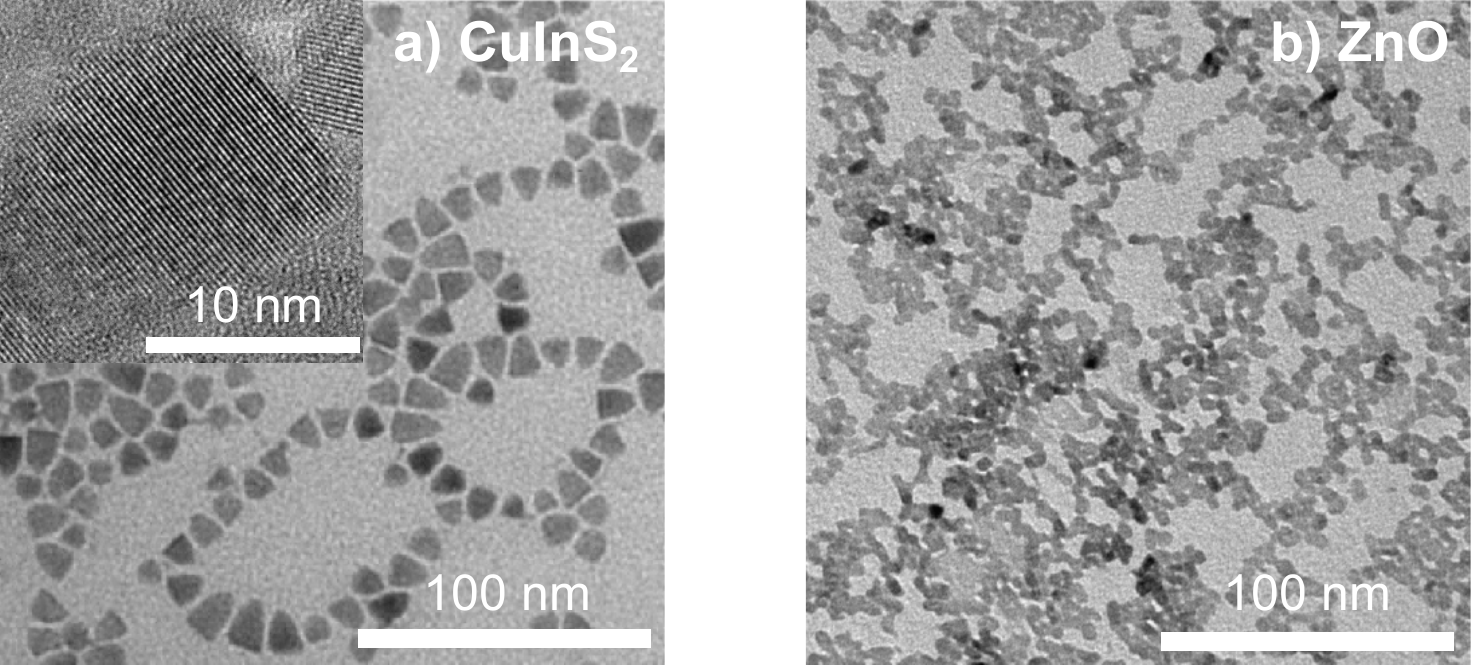}%
\caption{TEM images of the (a) \ce{CuInS2} and (b) ZnO NCs used within this study. The inset shows a HRTEM image of a single \ce{CuInS2} particle.}
\label{fig:S1}
\end{figure}

Figure \ref{fig:S1}a shows transmission electron microscopy (TEM) images of the pyramidally shaped \ce{CuInS2} colloidal nanocrystals (NCs) after ligand exchange with 1-hexanethiol. The edge length of the \ce{CuInS2} particles used within this study ranged from 12 to \unit[18]{nm}. The inset shows a high-resolution TEM images of a single \ce{CuInS2} particle. Figure \ref{fig:S1}b depicts the utilized ZnO NCs with a quasi-spherical shape and a diameter of $\sim$\unit[5]{nm}.

\clearpage

\section{Champion Device Characteristics}
\begin{figure}[h!]
\includegraphics[width=0.65\textwidth]{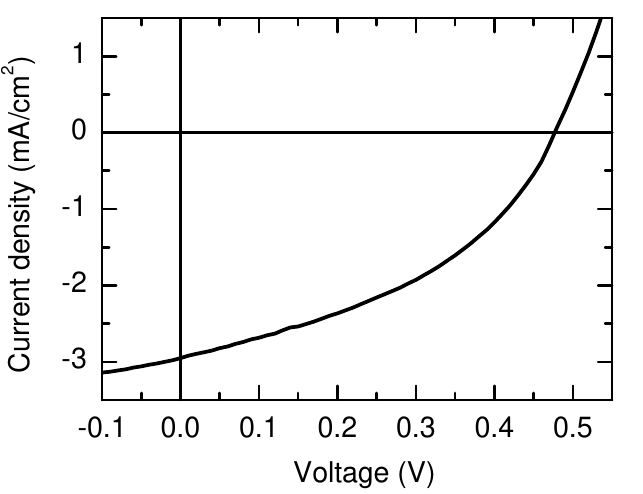}%
\caption{$J$--$V$ curve of the champion device obtained within this study.}
 \label{fig:S2}
\end{figure}

The $J$--$V$ curve of the best performing ITO\slash{}PEDOT:PSS\slash{}\ce{CuInS2}\slash{}ZnO\slash{}Al solar cell within the present study is displayed in Figure \ref{fig:S2}. That specific device had a ZnO layer thickness of \unit[60]{nm} and exhibited an open-circuit voltage of \unit[0.48]{V}, a short-circuit current density of \unit[3.0]{mA/cm$^2$}, a fill factor of 41\%, and a power conversion efficiency of 0.6\%.

\clearpage

\section{Determination of Optical Constants}
\begin{figure}[h!]
\includegraphics[width=\textwidth]{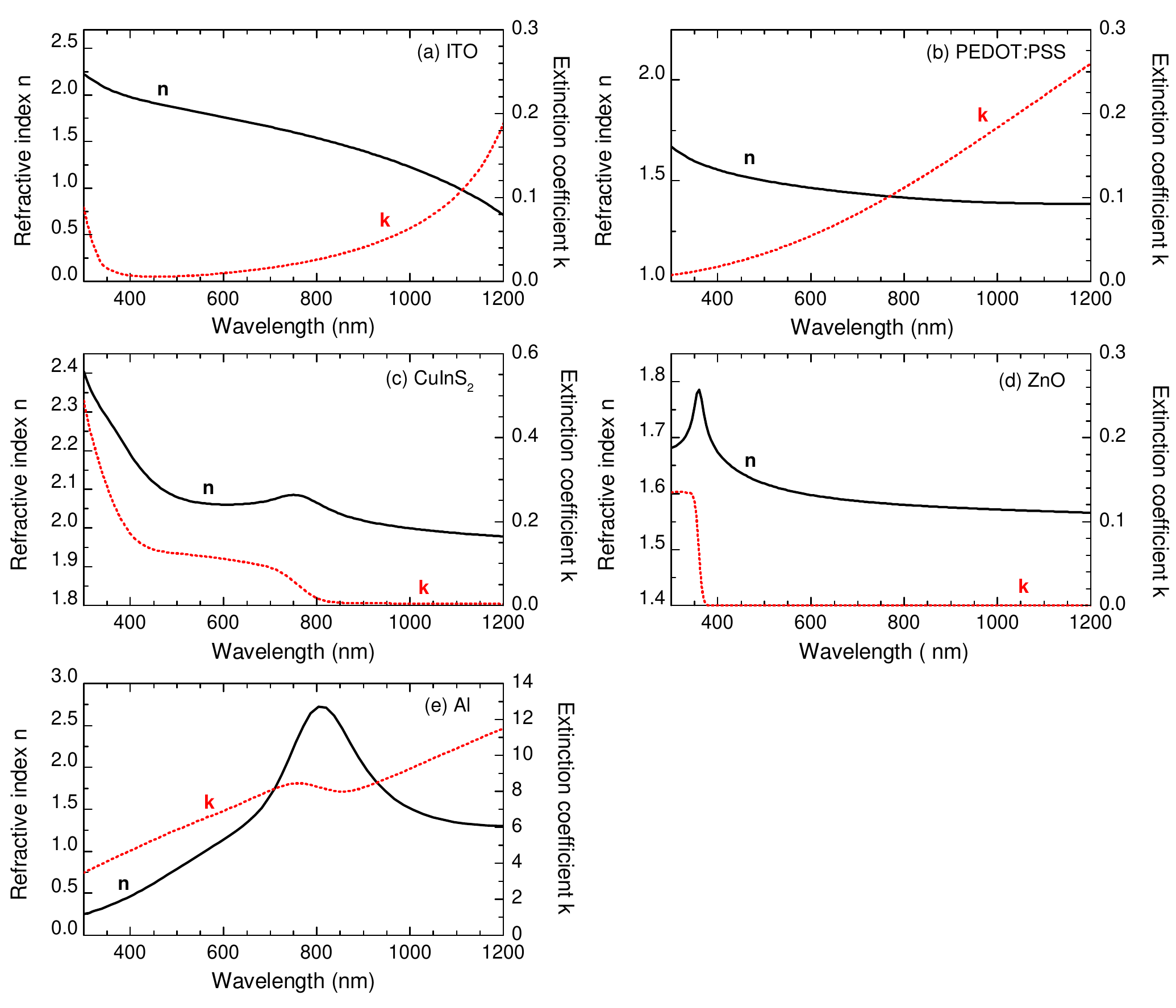}%
\caption{Spectral dependence of the refractive index ($n$) and extinction coefficient ($k$) for (a) ITO, (b) PEDOT:PSS, (c) CIS NCs, (d) ZnO NCs, and (e) Al, respectively, that were used within the optical simulations.}
\label{fig:S3}
\end{figure}

Figure \ref{fig:S3} shows the optical constants (refractive index $n$ and extinction coefficient $k$) that were used as input parameters to model the optical characteristics of our solar cells. The depicted values for ITO, PEDOT:PSS, \ce{CuInS2}, and ZnO were obtained by means of variable angle spectroscopic ellipsometry (VASE), as it is described in detail below. In case of Al, the data was taken from literature.\cite{Rakic1998}

For the VASE measurements, thin films of the respective materials were deposited under the same conditions as for the solar cell fabrication. Either float glass or Si wafers (with a \unit[2]{nm} native oxide layer) were used as substrate. The dispersion parameters of the substrate materials were carefully determined prior to the other measurements. VASE data was acquired using a rotating analyzer ellipsometer (J. A. Woollam VASE) in ambient atmosphere. The measurements were carried out under at least 3 incident angles ranging from 55$\circ$ to 75$\circ$ and for wavelengths between 280 and \unit[1700]{nm} (\unit[5]{nm} step size). The supplied software WVASE32 (J. A. Woollam) was used for the data analysis. First, we determined the film thickness by applying a simple Cauchy dispersion model to the spectral region where the films were transparent ($k = 0$). Afterward, the thickness was fixed and the ellipsometric data was modeled over the full wavelength range by means of a Kramers--Kronig consistent generalized oscillator model (GOM) consisting of typical parameterized oscillator functions. The mean square error (MSE) was used to quantify the difference between the experimental and modeled data. In the following, we present details regarding the sample preparation and data analysis for each of the investigated materials.

\subsection{Indium tin oxide (ITO)}
The optical constants of the commercially purchased ITO-covered glass substrates (Pr\"azisions Glas \& Optik GmbH, Germany) were determined according to the procedure described in our previous publication.\cite{Wilken2015}

\subsection{PEDOT:PSS}
The VASE measurements were performed on a spin-coated film of PEDOT:PSS (Clevios P VP AI 4083; thickness \unit[30]{nm}) on glass. Experimental data was modeled with a GOM consisting of one Tauc--Lorentz and one Lorentz oscillator. Finally, we obtained a MSE of 3.7. The optical constants were cross-checked by simulating the spectral reflectance of a PEDOT:PSS film on glass with the as-obtained optical constants and comparison with an independent reflectance measurement.

\subsection{\ce{CuInS2} NCs}
\begin{figure}[h!]
\includegraphics[width=0.6\textwidth]{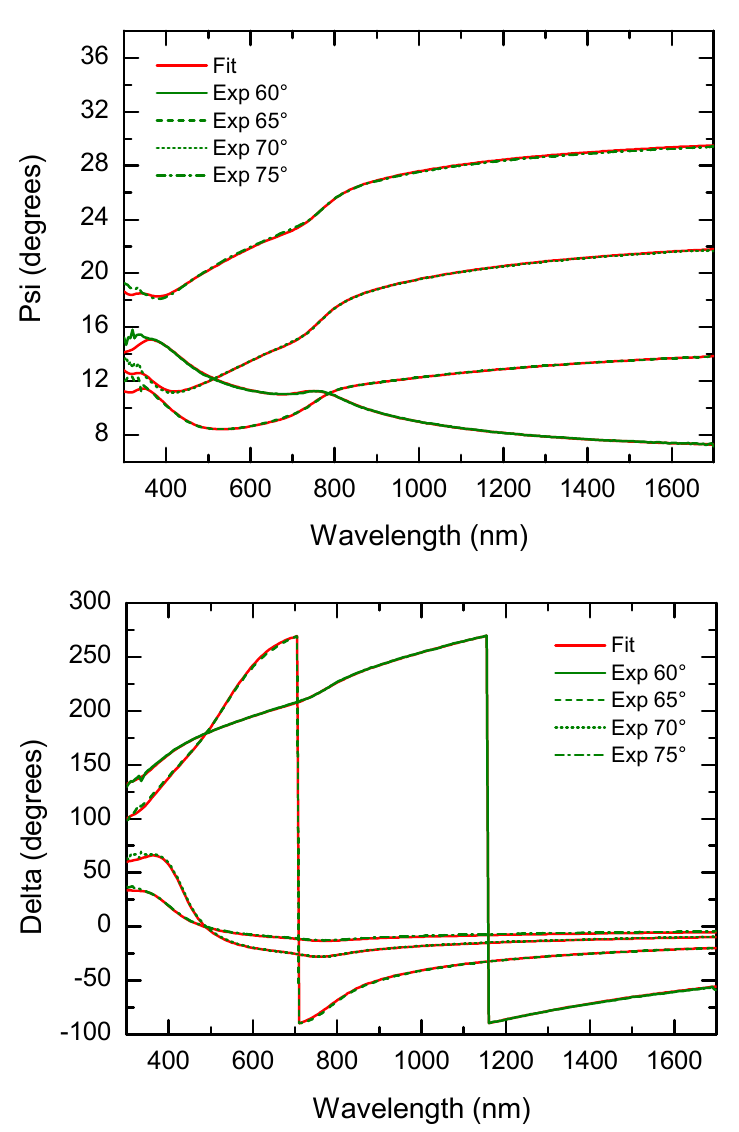}%
\caption{Comparison of the measured (green) and modeled (red) ellipsometric quantities $\Psi$ and $\Delta$ for a \unit[80]{nm} thick film of \ce{CuInS2} NCs on glass.}
\label{fig:S4}
\end{figure}

The VASE measurements were performed on a \unit[80]{nm} thick spin-coated film of \ce{CuInS2} nanoparticles on glass. For the data analysis, we employed a GOM consisting of three Herzinger--Johs parametric oscillator model functions (Psemi-M0, Psemi-M2, and Psemi-Tri). The finally obtained MSE was 2.3. Figure \ref{fig:S4} visualizes the excellent agreement between the modeled and measured ellipsometric quantities $\Psi$ and $\Delta$, indicating the high accuracy of the employed model. Moreover, the as-obtained optical constants were cross-checked with reflectance measurements, as well as comparison of the modeled absorption coefficient $\alpha =  4\pi k/\lambda$ with the previously reported one,\cite{Scheunemann2013} which was obtained on the basis of the spectral transmittance and reflectance of a \ce{CuInS2} film on glass. In both cases, we observed a high correlation.

\subsection{ZnO NCs} The VASE measurements were performed on a \unit[20]{nm} thick spin-coated film of ZnO nanoparticles on Si. The experimental data was modeled with a GOM consisting of one Herzinger--Johs parametric oscillator model function (Psemi-M0) and two Gaussian oscillators. The finally obtained MSE was 0.8. 

\subsection{Aluminum}
\begin{figure}[h!]
\includegraphics[width=0.6\textwidth]{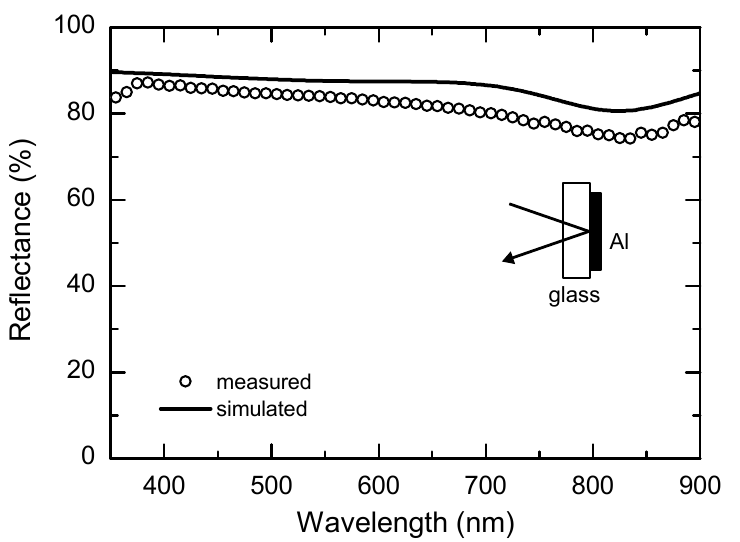}%
\caption{Comparison of the measured (circles) and simulated (line) spectral reflectance of a glass/Al/air three-layer system to validate the optical constants for Al taken from literature.}
 \label{fig:S5}
\end{figure}

Because of the oxide layer formation at the surface, the optical constants of Al could not be determined in a reliable manner by means of VASE measurements. Instead of that, we used literature data provided from Ref.~\citenum{Rakic1998}. These values were validated with the measured reflectance of a thermally evaporated Al film (thickness \unit[120]{nm}) on glass. Because of the fact that the oxidation of the Al layer only took place at its outer surface, which was in direct contact with the ambient atmosphere, but not at that one protected by the substrate, the sample was excited from the glass side. In Figure \ref{fig:S5}, we compare the result from this measurement to the reflectance one would expect on the basis of the optical constants taken from literature. For the latter, we used transfer-matrix method calculations under the assumption of a glass/Al/air three-layer system.

\clearpage

\section{Variation of \ce{CuInS2} Layer Thickness}
\begin{figure}[h!]
\includegraphics[width=0.6\textwidth]{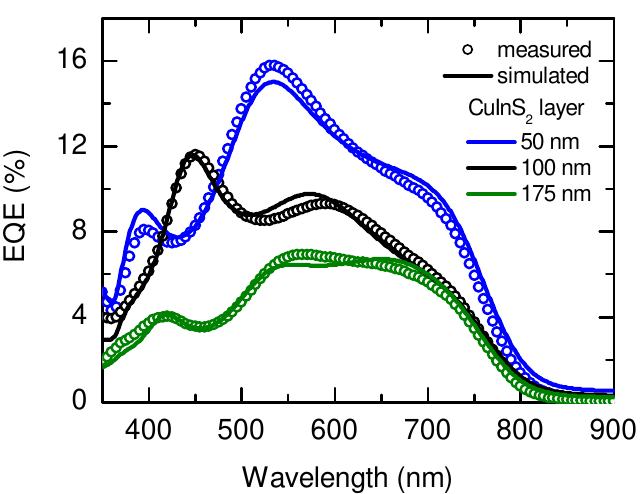}
\caption{Comparison of the measured (circles) and reconstructed EQE spectra after the optimization procedure (lines) for different thicknesses of the \ce{CuInS2} layer on an absolute scale.  Both the experimental and simulated data presented here is an average over several devices.}
\label{fig:S6}
\end{figure}

\begin{figure}[h!]
\includegraphics[width=0.6\textwidth]{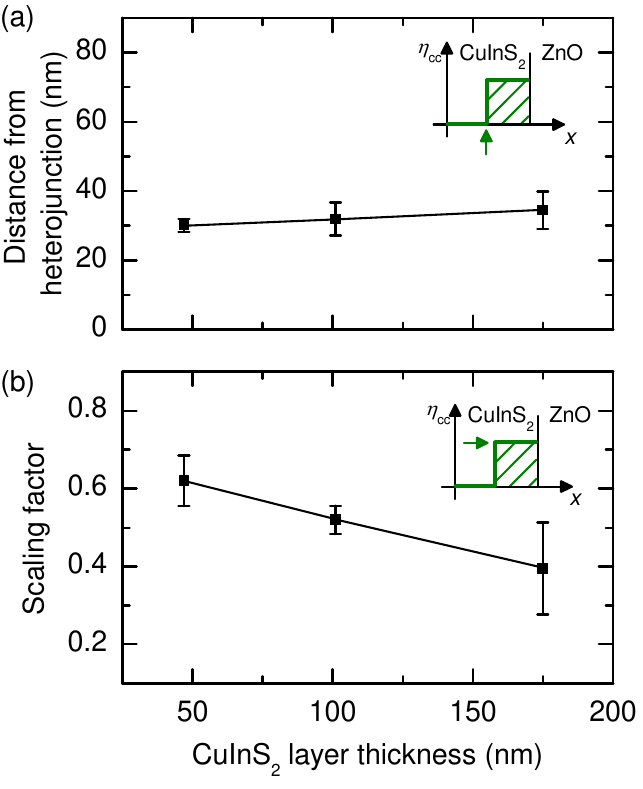}
\caption{Results for the optimization of $\eta_\text{cc}(x)$ in dependence on the \ce{CuInS2} layer thickness. (a) Position of the step edge in terms of the distance from the \ce{CuInS2}/ZnO interface. (b) Scaling factor related to the height of the step function. The data presented is an average of several devices with the corresponding standard deviation (error bars).}
\label{fig:S7}
\end{figure}

In Figure \ref{fig:S6}, the reconstructed EQE spectra, which were calculated from the simulated absorption profiles and the step-like charge collection probability function, are displayed in comparison to the measured ones for variable thickness of the \ce{CuInS2} layer. The reconstructed spectra shown here were obtained after optimizing the step function in terms of minimizing the mean square error. This optimization procedure was performed individually for several devices, and the extracted results for the step position and the step height are shown in Figure \ref{fig:S7}.

\clearpage

\section{Voltage-Bias Dependent EQE}
\begin{figure}[h!]
\includegraphics[width=\textwidth]{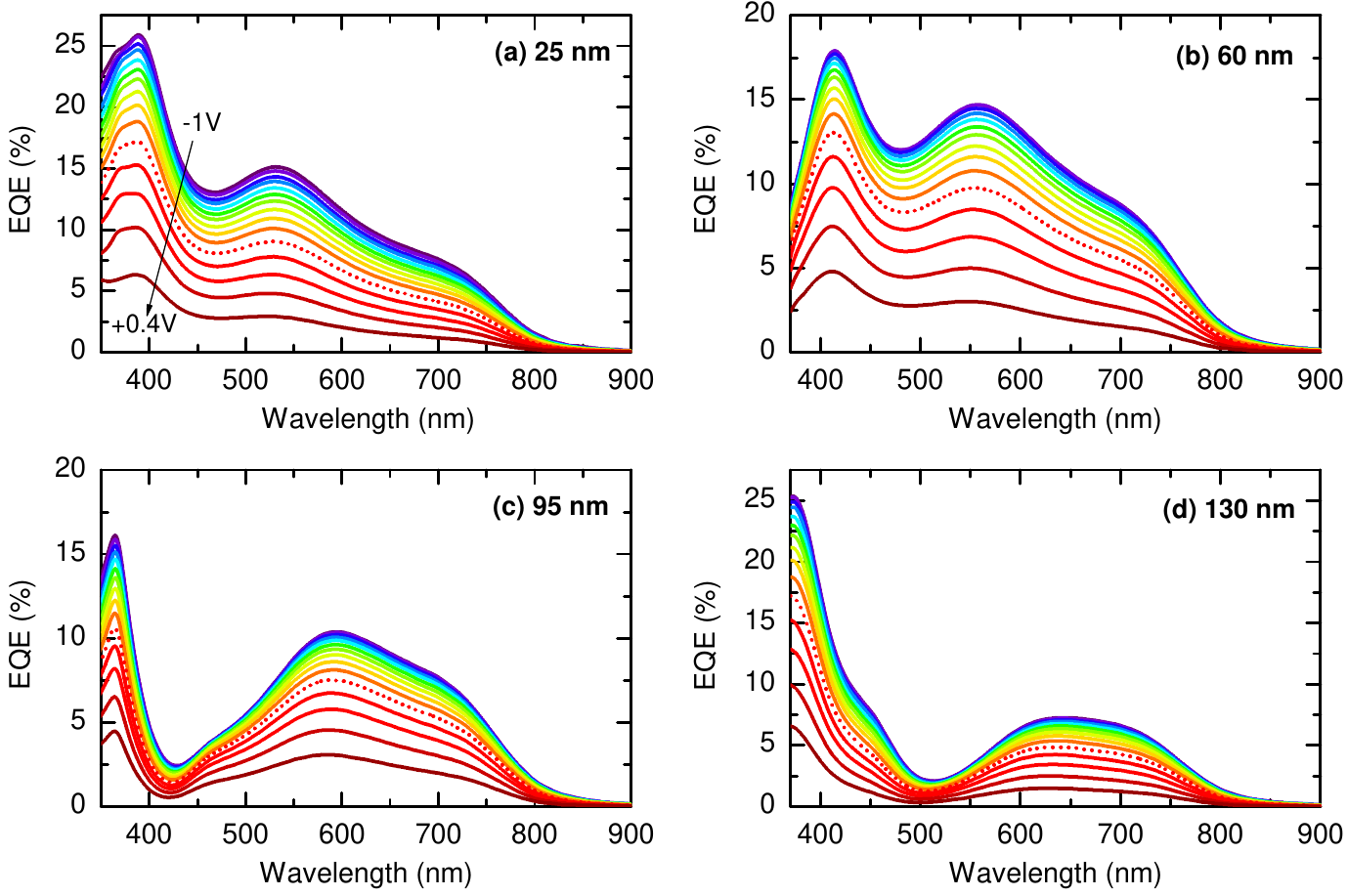}
\caption{Voltage-bias dependent EQE spectra of exemplary devices with ZnO layer thicknesses of (a) \unit[25]{nm}, (b) \unit[60]{nm}, (c) \unit[95]{nm}, and (d) \unit[130]{nm}. The bias voltage was varied from $-1$ to \unit[0.4]{V} in \unit[0.1]{V} increments. Note that the white light bias was still present during the measurements. Dotted lines denote the EQE obtained under short-circuit conditions.}
\label{fig:S8}
\end{figure}

Figure \ref{fig:S8} shows EQE spectra which were measured with an additionally applied bias voltage. The external voltage was provided by a low-noise current preamplifier (Stanford Research Systems SR570) connected in series to the sample. The bias voltage was varied from $-1$ to \unit[0.4]{V} in \unit[0.1]{V} increments.

\clearpage

\bibliography{references}